\documentclass[11pt]{article}
\usepackage{graphicx}
\usepackage{dcolumn}     % Align table columns on decimal point
\usepackage{bm}

\long\def\inst#1{\par\nobreak\kern 4pt\nobreak
    {\it #1}\par\vskip 10pt plus 3pt minus 3pt}

\begin{document}
{\pagestyle{empty}

\begin{flushright}
MUC-CONF-PHYSICS-299\\
%\babar-CONF-\BABARPubYear/\BABARConfNumber \\
%\babar-PUB-\BABARPubYear/\BABARPubNumber \\
%SLAC-PUB-\SLACPubNumber \\
%hep-ex/\LANLNumber \\
November 2004 \\
\end{flushright}

\par\vskip 5cm

% Title of the paper
\begin{center}
\Large \bf \boldmath 
Resolution of Nearly Mass Degenerate Higgs Bosons\\ and 
Production of Black Hole Systems of Known Mass\\ at a Muon Collider
\end{center}
\bigskip

\begin{center}
R.\ Godang\footnote{Correspondent: godang@phy.olemiss.edu}, 
S.\ Bracker, M.\ Cavagli$\rm\grave{a}$, L.\ Cremaldi and D.\ Summers\\
University of Mississippi-Oxford, University, MS 38677\\
\vspace*{0.5cm}
D.\ Cline\\
University of California-Los Angeles, Los Angeles, CA 90095
\end{center}
\bigskip \bigskip

% Abstract
\begin{center}
\large
\end{center}
The direct s-channel coupling to Higgs bosons is 40000 times greater for muons
than electrons; the coupling goes as mass squared. High precision scanning of
the lighter $h^0$ and  the higher mass $H^0$ and $A^0$ is thus possible with a
muon collider. The $H^0$ and $A^0$ are expected to be nearly mass degenerate
and to be CP even and odd, respectively. A muon collider could resolve the mass
degeneracy and make CP measurements.  The origin of CP violation in the $K^{0}$ and
$B^{0}$ meson systems might lie in the the $H^0$/$A^0$ Higgs bosons. 
If large extra dimensions exist, black holes with lifetimes  of $\sim 10^{-26}$ 
seconds could be created and observed via Hawking radiation at the LHC. 
Unlike proton or electron colliders, muon colliders can produce black hole 
systems of known mass. This opens the possibilities of measuring quantum remnants, 
gravitons as missing energy, and scanning production turn on. 
Proton colliders are hampered by parton distributions and CLIC by beamstrahlung. 
The ILC lacks the energy reach.

\vspace{2.0cm}

\begin{center}
Proceedings to the DPF 2004: Annual Meeting of the Division of Particles and Fields of APS\\
26 August-31 August 2004, Riverside, CA, USA
\end{center}

\newpage

} % end of pagestyle{empty}

% The body of the paper starts here

\section{Introduction}

A $\mu^{+}\mu^{-}$ collider with 4 TeV center-of-mass (CM) energy has unique
and prominent features to probe the Higgs boson physics and
produce microscopic black holes of known CM energy in super-Planckian 
events\,\cite{marco_modern03}. Muons emit 2 billion times less bremsstrahlung
radiation then electrons ($(m_{\mu}/m_{e})^4$).
This makes the muon collider\,\cite{muon_col03} very attractive: 
1) The possibility of precision measurements of
the Higgs and supersymmetric particles using rings; 
2) The possibility of the separation of a higher Higgs boson doublet by scanning
s-channel production; 3) Production of black hole (BH) systems (initially radiated 
gravitons plus BH) of known mass without the limitations of parton distributions or 
beamstrahlung smearing\,\cite{azuelos}.

\section{Heavy Higgs Bosons}

The Higgs discovery will likely occur at Large Hadron Collider (LHC) experiments
such as the Compact Muon Solenoid (CMS)\,\cite{godang04}. The next critical
challenge in high-energy physics will be to differentiate between the Standard
Model (SM) Higgs boson and the Minimal Supersymmetric Standard Model (MSSM)
Higgs boson. In the context of MSSM, the important Higgs parameters are $m_{A^{0}}$
and $tan~\beta$, where $tan~\beta$ is the ratio of the vacuum expectation
values of the Higgs doublet. In the first-order approximation, the Higgs
decay width is small compared to the mass resolution for large $tan~\beta$.
Furthermore, the higher mass Higgs bosons, $H^0$/$A^0$, may be nearly mass degenerate.
The LHC and the ILC may fail to separate them.

\begin{figure}[!htb]
\begin{center}
\includegraphics[width=9cm]{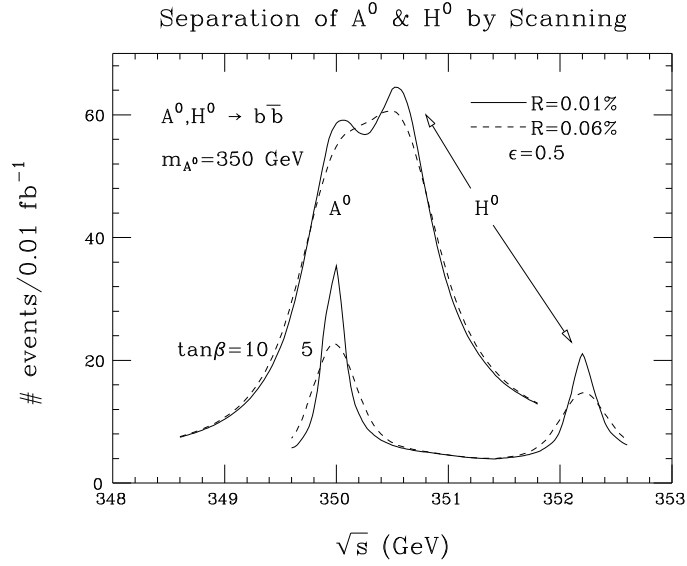}
\vspace*{0.1cm}
\caption{Separation of heavy Higgs bosons $H^0$ and $A^0$ of $b \bar b$ final
state event rate as a function of $\sqrt{s}$ for $m_{A^{0}}$ = 350 GeV
and $tan~\beta$ = 5 and 10. Courtesy of J.\ F.\ Gunion, Ref.~\cite{barger95}.}
\label{fig:AH}
\end{center}
\end{figure}

For the s-channel Higgs process of narrow resonances, energy resolution is 
an important consideration. A muon collider with sufficient energy resolution
(R = 0.01\% and/or 0.06\%) might be the only possible means for separating   
the two higher mass Higgs bosons, $H^0$ and $A^0$. The s-channel Higgs
resonance would be found by scanning in $\sqrt{s}$ using a small step
($\sigma$). Using sufficient energy resolution, the line shape of a
Breit-Wigner resonance and the Higgs width could be deduced. 
Figure~\ref{fig:AH} shows the $H^0$ and $A^0$ resonances for $tan~\beta$ = 5
and 10, including the $b \bar b$ continuum background\,\cite{barger95}. It is
clear the resonances are overlapping for the larger value of $tan~\beta$. With
R = 0.01\%, the two distinct resonance peaks are clearly visible, although, the
peaks are smeared out for R = 0.06\%.

\section{Black Holes Production}

BH production at colliders depends on the value of the fundamental Planck scale
$M_{Pl}$, which determines the energy where the gravitational interaction
becomes strong. If this fundamental energy scale is as low as the TeV scale,
BHs can be produced in future colliders such as the LHC, CLIC, and the muon
collider.  BH production at the LHC depends on parton distributions. 
BH production at the CLIC is smeared by beamstrahlung. 
These effects introduce extra unknowns into the physics of BH formation. 
Unlike the LHC and the CLIC, the muon collider could produce BH systems of 
known energy with a simple cross section.

The properties of BHs at the muon collider are completely determined by the BH
mass and angular momentum. In this paper we will restrict our discussion to BHs
with vanishing angular momentum for simplicity. In a scenario with $n$ large
flat extra dimensions, the cross section of BH formation at the muon collider
is roughly the geometrical cross section
\begin{equation}
\sigma_{\mu\mu\to BH}(E) = F 
\frac{1}{M_{Pl}^2}\left[\frac{8\Gamma\left(\frac{n+3}{2}\right)}{(2+n)}
\frac{\sqrt{s}}{M_{Pl}}\right]^\frac{2}{n+1},
\end{equation}
where $F$ is a form factor which depends on the model. The semiclassical
approach suggests that if the impact parameter is less than the Schwarzschild
radius corresponding to a BH with mass $M_{BH}=\sqrt{s}$, a BH is formed. 
After formation, the BH evaporates semiclassically in a time $\sim 10^{-26}\,s$
through emission of Hawking radiation\,\cite{hawking}. The BH emits SM fields on
the brane and gravitational quanta in the extra-dimensional bulk.  

\begin{figure}[ht]
\vspace*{-1.5cm}
\begin{minipage}{0.50\textwidth}
\includegraphics[width=9cm]{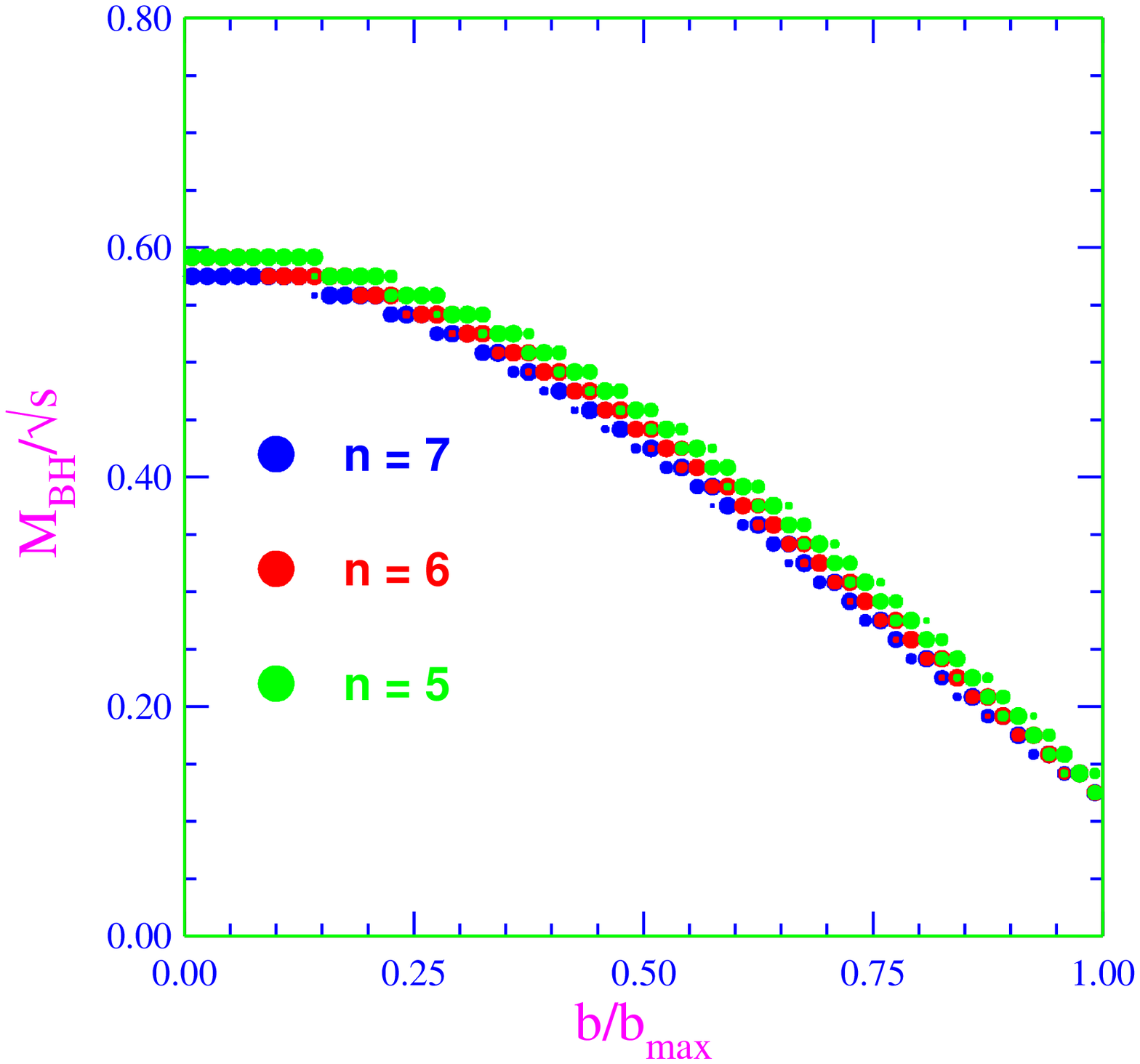}
\vspace{-20mm}
\caption{The correlation between the horizon mass, $M_{BH}/\sqrt s$, and
the impact parameter ratio, $b/b_{max}$, for $n=5,\dots 7$ large extra 
dimensions.}
\label{fig:formation}
\end{minipage}\hfill 
\begin{minipage}{0.46\textwidth}
\vspace*{1.4cm}
\includegraphics[width=8cm]{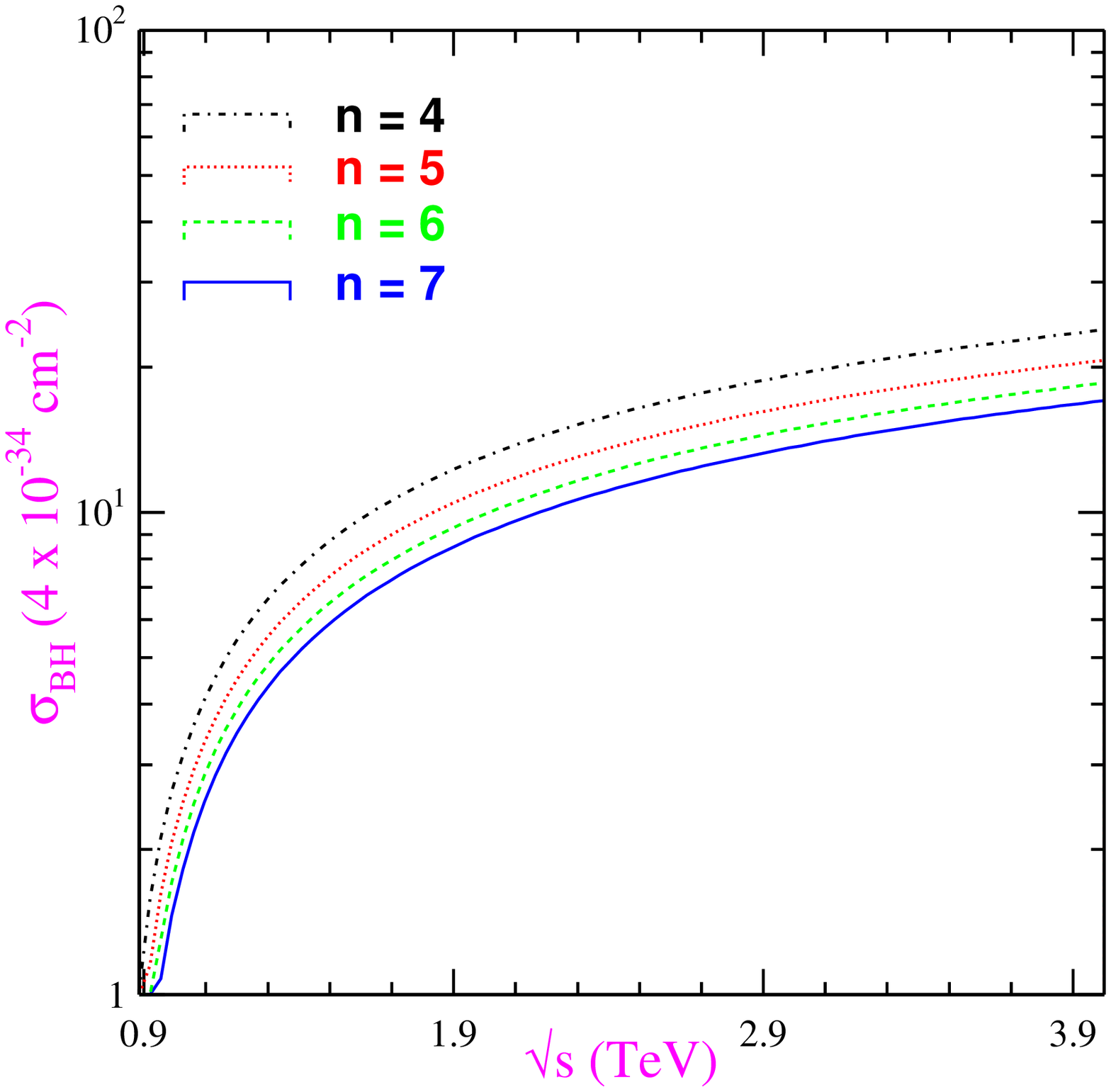}
\vspace{-1.5mm}
\caption{The BH cross section produced at a muon collider
as a function of the CM energy for $n=4,\dots 7$ large extra dimensions.}
\label{fig:cross-section}
\end{minipage}
\end{figure}

For totally elastic processes, the mass of the BH is equal to the CM energy
of the collision. For inelastic collisions, some of the CM energy is not
trapped into the BH. An indicator of the energy trapped by the horizon is given
in Ref.~\cite{nambu}. The trapped energy monotonically decreases with the
impact parameter from a maximum value of about 60\% of the CM energy (head-on
collision) to zero (maximum impact parameter $b_{max}$).
Figure~\ref{fig:formation} shows the correlation between the $M_{BH}/\sqrt s$
and the impact parameter ratio $b/b_{max}$ for different numbers of large 
extra dimensions. The energy trapped by the horizon decreases with more dimensions. 

The BH cross section for a muon collider is shown in Fig.~\ref{fig:cross-section} 
as a function of the collider CM energy. The typical BH cross section at a 4 TeV
muon collider is of order of a few $nb$. By using the lower bound of the
integrated luminosity, ${\cal{L}}_{\mu^{+}\mu^{-}}=  10^{33}\,(cm^{-2}s^{-1})$,
the BH production rate is about 7 BHs/s.

\section{Black Hole Experimental Signatures}

BH experimental signatures at the muon collider are large production cross section, 
events with large missing total/transverse energy ($E_{T}$), and high $E_{T}$ jets. 
If the BH leaves no remnant, the expected total missing energy for a 4 TeV 
collision is about 2.7 TeV. 
This missing energy is essentially due to gravitons propagating in the
extra-dimensional bulk, plus a few gravitons and neutrinos on the brane.
Assuming an isotropic distribution of emitted gravitons in the bulk and a
four-body final decay for the BH, the missing transverse energy due to the transverse
momentum of the gravitons and neutrinos on the brane is about 190 GeV. The
typical hadron to photon (lepton) ratio of BH decay is about 60:1 (7:1). 

\section{Summary and Acknowledgments}

A muon collider is fundamental to study the direct s-channel Higgs boson and 
differentiate between the $A^{0}$ and $H^{0}$. 
BH production at a 4 TeV CM energy muon collider does not suffer from beamstrahlung 
smearing and unknown factors of parton production. Unlike proton or electron colliders, muon
colliders can produce black hole systems of known mass. These unique features
open the possibilities of measuring BH quantum remnants, bulk graviton emission,
and Hawking radiation on the brane.

The authors would like to thank the Muon Collider Collaboration.
This work was supported in part by the U.S. Department of Energy 
contract DE-FG05-91ER40622.

%%CITATION = IMPAE,A18,1843;%%
%%CITATION = APCPC,352,4;%%
%%CITATION = APCPC,352,5;%%
%%CITATION = APCPC,352,6;%%
%%CITATION = SJPNA,12,223;%%
%%CITATION = PLACB,14,75;%%
%%CITATION = APCPC,156,201;%%
%%CITATION = APCPC,352,204;%%
%%CITATION = NUIMA,A350,27;%%
%%CITATION = NUPHZ,51A,61;%%
%%CITATION = JPHGB,G29,1577;%%
%%CITATION = PHYS-ICS 9911009;%%
%%CITATION = PRSTA,2,081001;%%
%%CITATION = PRSTA,6,081001;%%
%%CITATION = PHRVA,D58,013005;%%
%%CITATION = NUIMA,A532,470;%%
%%CITATION = HEP-PH 0204031;%%
%%CITATION = PRLTA,75,1462;%%
%%CITATION = HEP-PH 9804358;%%
%%CITATION = CMPHA,43,199;%%
%%CITATION = PHRVA,D67,024009;%%

\end{document}